\documentclass[12pt]{article}
\usepackage{floatfig,multicol}

\def\et{et al.}

\begin{document}
\initfloatingfigs

\begin{center}
{\LARGE 
{\bf 
On the way to global phase-delay astrometry
}
}\\[3mm]

{\large 
E. Ros$^1$, J.M. Marcaide$^2$, J.C. Guirado$^2$, 
M.A. P\'erez-Torres$^3$
}\\[3mm]

{\footnotesize 
{\it
$^1$ Max-Planck-Institut f\"ur Radioastronomie, Bonn, Germany

$^2$ Dep.\ Astronomia i Astrof\'{\i}sica, Universitat de Val\`encia, 
Burjassot (Valencia), Spain

$^3$ Istituto di Radioastronomia/CNR, Bologna, Italy
}
}
\end{center}

\begin{small}
\begin{quote}
The use of the phase-delay improves substantially the
accuracy obtained in VLBI astrometry 
with respect to the group-delay observable.  Recently, 
Ros et al.\ (1999) have extended the related
phase-connection technique to a triangle of 
radio sources with relative separations up to 6.8$^\circ$
(the S5 sources BL\,1803+784/QSO\,1928+738/BL\,2007+777).  
This technique has 
also been extended for separations up to 15$^\circ$ in the studies
of the pair of S5 radio sources QSO\,1150+812/BL\,1803+784 
(P\'erez-Torres et al.\ 2000).  
We are carrying out a 
long-term astrometric programme at 8.4, 15, and 43\,GHz to determine the 
absolute kinematics of radio source components in the 
13 members of the complete S5 polar cap sample.
For each epoch, all the radio sources are observed at the same frequency
for 24\,hours in total.
Bootstrapping 
techniques are used to phase-connect the data jointly for the 
13 objects throughout the observations.  An accurate 
registration of the maps of the radio sources at different 
frequencies will allow a study
of jet components with unprecedented precision and provide
spectral information.
This observing scheme could be extended in the future to other regions
of the sky and eventually lead to global phase-delay astrometry.
\end{quote}
\end{small}

\section{Introduction}

The improvement by four orders of magnitude in wide-angle
astrometric measurements in the last 30 years, provided mainly by
Very Long Baseline Interferometry (VLBI),
have resulted in
a better definition of the fundamental astronomical reference
frame (Johnston \& de Vegt 1999),
the International Celestial Reference Frame (ICRF), 
based on the radio positions of 212 compact extragalactic radio sources.
Up to the present, the group delay VLBI observable
has been regularly used for such a task.  The use of the more
precise 
phase delay  observable should provide an immediate improvement
in accuracy. 
A phase-connection process is needed (Shapiro \et\ 1979)
to overcome the inherent $2\pi$ ambiguity of the phases.
When used with radio source pairs, the phase delay
becomes the most accurate observable in astrometry.
The phase-delay astrometric technique has been applied to different 
pairs of radio sources with separations
from minutes to some degrees in the sky
(among others, 3C\,345/NRAO512, Shapiro \et\ (1979), Bartel \et\ (1986);
1038+528\,A/B, Marcaide \et\ (1983, 1995) and Rioja \et\ (1997);
4C\,39.25/0920+390, Guirado \et\ (1995b); 
QSO\,1928+738/BL\,2007+777, Guirado \et\ (1995a, 1998); 
3C\,395/3C\,382, Lara \et\ (1996)).

A significant technical problem in the 
astrometric data reduction is the modelling and removal of
the ionospheric effect.   Since the latter is a frequency-dependent
effect, the removal was done in the past by observing at two frequencies.  
The Global Positioning System (GPS) allows now to
obtain ionospheric data of unprecedented accuracy.  
Ros \et\ (2000a) and P\'erez-Torres \et\ (2000) used
GPS data to successfully remove the ionospheric effect.  Such an approach
can be applied to any VLBI astrometric observations, without
the need for dual-frequency observations.

Many compact radio sources show variable structures on scales 
larger than the accuracy of their position estimates.
If the source structure is ignored, 
the group-delay astrometric results may incorrectly be interpreted 
as a proper motion of the radio sources.
Because the radio sources, with typical redshifts of 1.0, are very distant,
proper motions should not be detectable.
It is true that most of the emission is generally produced 
in the core (brightest feature) of the source, but for low 
frequencies, due to opacity effects, this core may not be 
close to the central 
engine of the source (mass center of the system, which 
sould be stationary).  Until the efforts 
of Fey \et\ (1996) and Fey \& Charlot (1997, 2000) the effect 
of source structure had been ignored in the group-delay based 
reference system maintenance.  These authors have initiated
a programme to image the radio reference frame sources and define
a source ``structure index" to provide a quantitative estimate
of the possible astrometric quality of them.
In contrast, the phase-delay analysis from
the early studies of Shapiro \et\ (1979) 
takes into account
the effect of the structure and to date, the
phase-delay astrometry is the only rigorous way to
register radio maps with precisions of 100\,$\mu$as or better.  As mentioned
above, this is the technique that we use in our astrometric
data analysis.

\section{Phase-delay astrometry of S5 polar cap sample sources}

Eckart \et\ (1986,1987) selected 13 radio sources from 
the MPIfR-NRAO 5\,GHz survey (abbreviated as S5, 
K\"uhr \et\ 1981) with declinations
over 70$^\circ$, flat spectral indices and flux densities over
1\,Jy at $\lambda6$\,cm.
These sources, given in Table \ref{table:s5} 
(IERS coordinates, optical
magnitude and redshift) constitute
the S5 polar cap sample. 
The sky distribution of the members of the sample,
with their angular separations indicated, is given in Fig.\ \ref{fig:poldiag}.  

\begin{table}[htb]
\caption{S5 polar cap sample.
\label{table:s5}} 
\vspace{4pt}
\begin{footnotesize}
\centerline{\begin{tabular}{rrrcc} 
\hline
\multicolumn{1}{c}{Name}&  \multicolumn{1}{c}{R.A.}& \multicolumn{1}{c}{Dec} & $V$ & $z$ \\ \hline
QSO\,0016+731 &  0$^h$19$^m$45$\rlap{.}^s$78645& 73$^\circ$27$'$30$\rlap{.}''$0175 & 18.0 & 1.781 \\ 
QSO\,0153+744 &  1$^h$57$^m$34$\rlap{.}^s$96500& 74$^\circ$42$'$43$\rlap{.}''$2305 & 16.0 & 2.338 \\ 
QSO\,0212+735 &  2$^h$17$^m$30$\rlap{.}^s$81337& 73$^\circ$49$'$32$\rlap{.}''$6218 & 19.0 & 2.387 \\ 
 BL\,0454+844 &  5$^h$ 8$^m$42$\rlap{.}^s$36340& 84$^\circ$32$'$ 4$\rlap{.}''$5438 & 16.5 & 0.112 \\ 
QSO\,0615+820 &  6$^h$26$^m$ 3$\rlap{.}^s$00612& 82$^\circ$ 2$'$25$\rlap{.}''$5680 & 17.5 & 0.710 \\ 
 BL\,0716+714 &  7$^h$21$^m$53$\rlap{.}^s$44848& 71$^\circ$20$'$36$\rlap{.}''$3634 & 14.2 & --\\ 
QSO\,0836+710 &  8$^h$41$^m$24$\rlap{.}^s$36529& 70$^\circ$53$'$42$\rlap{.}''$1735 & 16.5 & 2.172 \\ 
QSO\,1039+811 & 10$^h$44$^m$23$\rlap{.}^s$06255& 80$^\circ$54$'$39$\rlap{.}''$4430 & 16.5 & 1.254 \\ 
QSO\,1150+812 & 11$^h$53$^m$12$\rlap{.}^s$49923& 80$^\circ$58$'$29$\rlap{.}''$1546 & 18.5 & 1.250\\ 
 BL\,1749+701 & 17$^h$48$^m$32$\rlap{.}^s$84022& 70$^\circ$ 5$'$50$\rlap{.}''$7687 & 16.5 & 0.770 \\ 
 BL\,1803+784 & 18$^h$ 0$^m$45$\rlap{.}^s$68393& 78$^\circ$28$'$ 4$\rlap{.}''$0184 & 16.4 & 0.684\\ 
QSO\,1928+738 & 19$^h$27$^m$48$\rlap{.}^s$49521& 73$^\circ$58$'$ 1$\rlap{.}''$5699 & 15.5 & 0.302 \\ 
 BL\,2007+777 & 20$^h$ 5$^m$30$\rlap{.}^s$99855& 77$^\circ$52$'$43$\rlap{.}''$2478 & 16.5 & 0.342 \\ 
\hline
\end{tabular}}
\end{footnotesize}
\end{table}

Guirado \et\ (1995a, 1998)
studied astrometrically two sources of this sample at
2.3, 5, and 8.4\,GHz, QSO\,1928+738 and BL\,2007+777.  
Ros \et\ (1999) added a new source to this pair, BL\,1803+784,
studied the trio at 2.3 and 8.4\,GHz, and determined the
relative positions with $\sim130\,\mu$as
precisions at 8.4\,GHz.  The phase connection technique was
thus extended to angular separations of 6.8$^\circ$.
Later, P\'erez-Torres \et\ (2000) phase-connected data on the pair 
QSO\,1150+812/BL\,1803+784, separated
by 14.9$^\circ$.  The situation now allows us to design a strategy
to extend the phase-delay astrometric technique to larger
sets of radio sources. The S5 polar cap sample, 
given the high flux densities of the radio sources and
their situation on the sky, is optimal for carrying out
phase-delay astrometry simultaneously for the complete
sample.

\begin{floatingfigure}{280pt}
\vspace{280pt}
\includegraphics{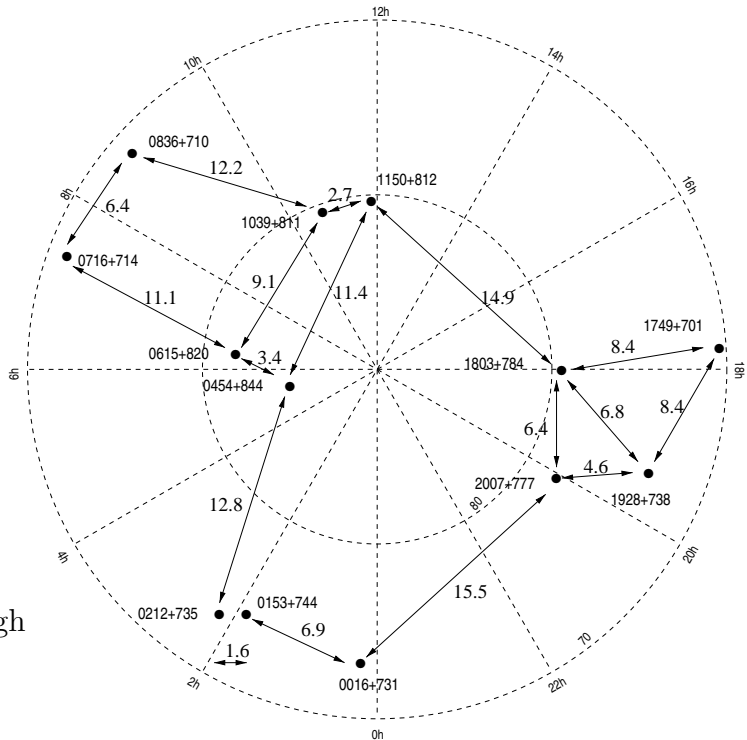}
\caption{\small Distribution of the S5 polar cap sample in the northern
sky, centered at the celestial north pole.  The black dots represent
the positions of radio sources, and the angular distances between
them are indicated in arc degrees with arrows.  \label{fig:poldiag} }
\end{floatingfigure}

All sources have neighbours at angular distances such
that the use of the phase-connection technique is possible 
throughout
a 24\,hr VLBA observing run.  Using the phase-delay data
from the 13 radio sources we should obtain their relative positions with
accuracies better than 0.1\,mil\-li\-arc\-sec\-onds (mas).

\begin{figure}[htbp]
\vspace{563pt}
\includegraphics{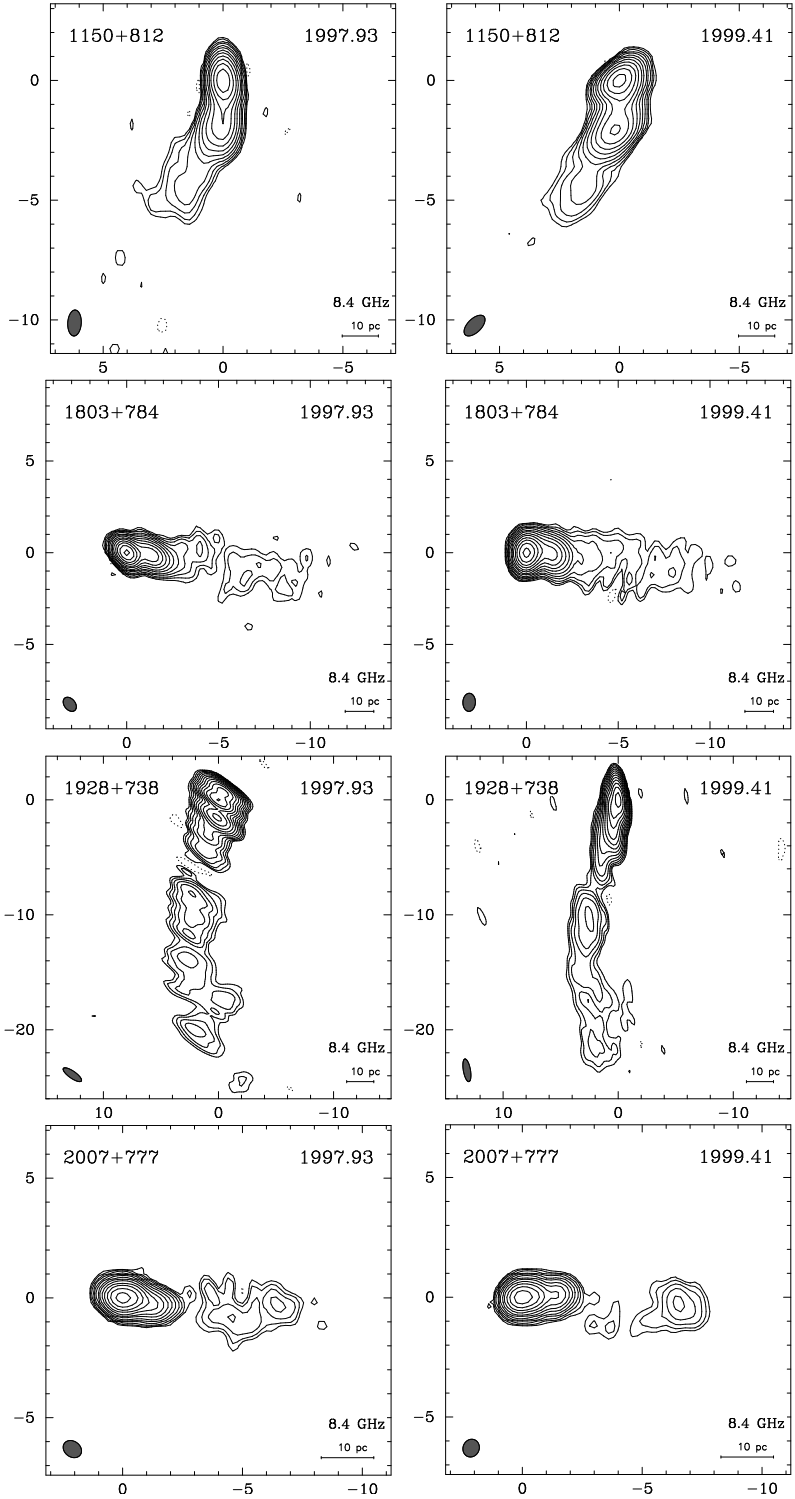}
\vspace{-92mm}
\hfill \parbox[b]{5.3cm}{\caption[]{
\small
VLBA images of QSO\,1150+812, BL\,1803+784, QSO\,1928+738,
and BL\,2007+777 from observations on 6 December 1997 (1997.93) and
28 May 1999 (1999.41).  Axes are relative $\alpha$ and $\delta$ in mas.
Minimum contour levels are of 1.0\,mJy/beam 
for QSO\,1150+812 and BL\,1803+784, 1.4\,mJy/beam for QSO\,1928+738,
and 0.9\,mJy/beam for BL\,2007+777.
Note that the convolution beam of a given source in a given epoch substantially
differs from those of other sources and epochs, 
due to the different coverages of the
$(u,v)$ plane.
\label{fig:8maps}} 
}
\end{figure}

We have observed all the 13 members of the sample at 8.4\,GHz at
epochs 1997.93 and 1999.41, and at 15\,GHz in 1997.57 and 2000.46 using 
a phase-re\-fe\-ren\-ce scheme. In each of the 24-hour observations at 
each wavelength, each source was observed on average for
5\,hours, enough to produce high-quality hybrid maps.  In 
Fig.\ \ref{fig:8maps} we show the 8.4\,GHz hybrid
maps from those sources
of the S5 polar cap sample that were studied astrometrically
in the past ---Guirado \et\ (1995a, 1998, 2000), Ros \et\ (1999), 
P\'erez-Torres \et\ (2000).

\begin{figure}
\vspace{172pt}
\includegraphics{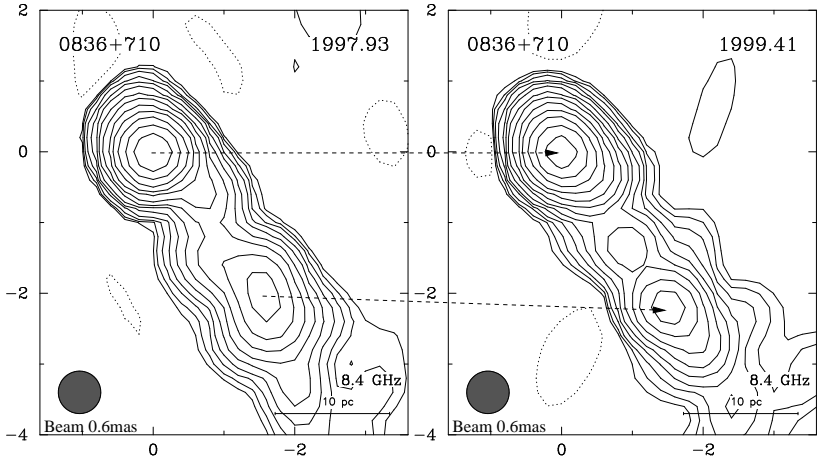}
\vspace{-47mm}
\hfill \parbox[b]{5.3cm}{\caption[]{
\small
VLBA images of QSO\,0836+710 convolved with a 0.6\,mas 
circular beam.  We observe the small structural changes in the 
inner part of the jet.  The 
dashed lines draw a tentative association between features from
one epoch to another.
\label{fig:0836-06}
}}
\end{figure}

It is possible to observe some morphological changes
just from the present mapping results at 8.4\,GHz,
as is the case
for the sources QSO\,0016+731 and QSO\,1928+738 (Ros \et\ 2000b).
As an example, we present 
in Fig.\ \ref{fig:0836-06}
the changes in the inner jet features of QSO\,0836+710 
after
convolving the {\sc clean}-components of the map 
with a 0.6\,mas circular beam to better compare the  images.
The evolution within 3\,mas of the core (P.A.\ $-140^\circ$) is evident
during the 1.5 years elapsed between the observations.

\begin{figure}[htb]
\vspace{172pt}
\includegraphics{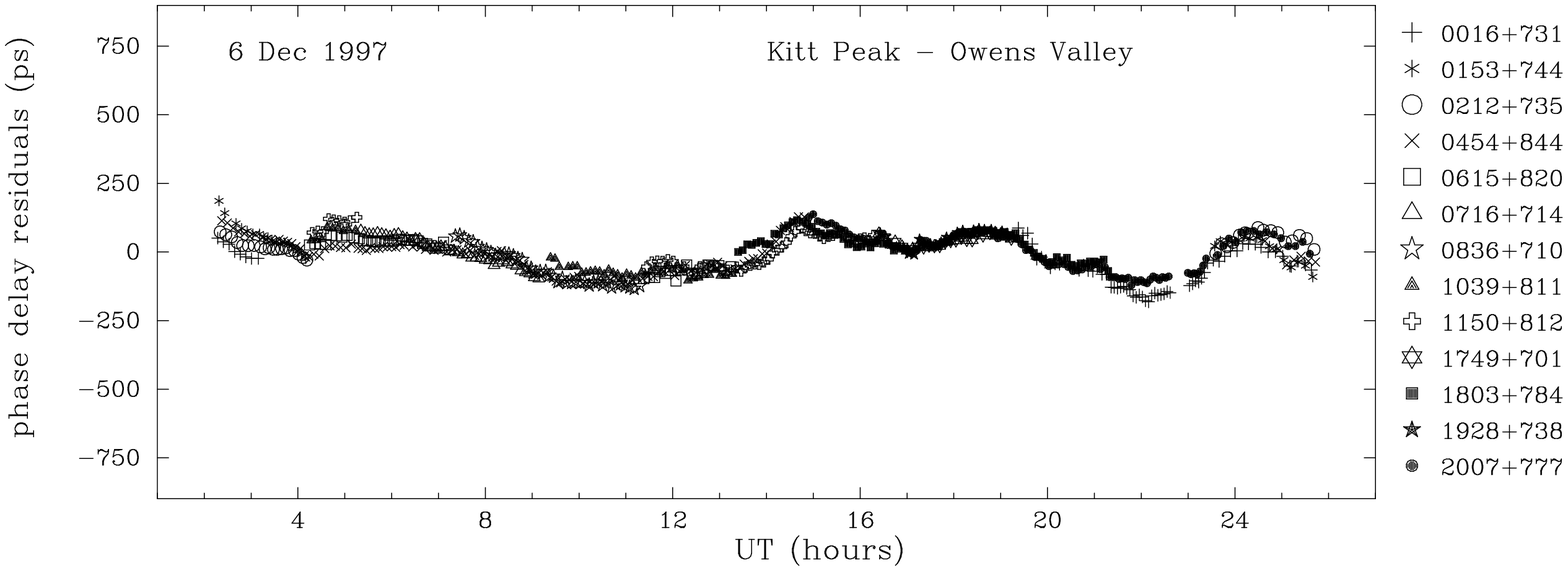}
\caption[] 
{\small Postfit residuals of the (undifferenced) phase-delays 
at 8.4\,GHz after a weighted 
least-squares analysis that estimates the relative separations between the
radio sources. One phase-cycle at 8.4\,GHz corresponds
to  120\,ps of phase-delay. 
Notice a similar trend for all sources.
These trends cancel out in the differences to yield a global residual rms of 
$\sim 30$\,ps.}
\label{fig:phas-conn}
\end{figure}

The astrometric data reduction of the S5 polar cap sample data is in progress.
We can report very interesting preliminary astrometric results from the 
observations of epoch 1997.93.  
The joint phase-delay analysis for all the 13 radio sources did
not present particular difficulties.  The quality of our
analysis is demonstrated by the low root-mean-square (30\,ps) post-fit
residuals shown in Fig.\ \ref{fig:phas-conn}.
The phase-connection process has also been eased by the
availability of an interactive, {\sc pgplot}-based software developed
in our group, which permits us to efficiently correct 
for phase-delay ambiguities. 
The expected average precision (including all systematic
errors) of the relative position determination of all radio sources
in the sample is of $\sim$80\,$\mu$as at 8.4\,GHz.  If extrapolated, these
results indicate that our observations at 15\,GHz should yield a precision of
$\sim$50\,$\mu$as.

Related to this work, Guirado \et\ (2000) have shown
the feasibility of precision differential phase-delay 
astrometry at 43\,GHz ($\lambda7$\,mm)
by studying the pair of S5 radio sources QSO\,1928+738 
and  BL\,2007+777, separated by 
$\sim 5^\circ$. 
For their results, the root-mean-square of the postfit
residual delays is $\sim 2$ ps, or equivalently $\sim 30^\circ$ 
of phase at this frequency.
There can be little doubt that the entire S5 sample can thus be studied at 
43\,GHz,
following studies at 8.4 and 15\,GHz, and
the relative positions of all the sources determined with 
$\sim 20 \mu$as precision.  The VLBA has allocated time
to observe the S5 polar cap sample also at 43\,GHz in late 2000.
Thus, the observations at 8.4 and 15\,GHz will be complemented
with those at mm-wavelength.

\section{Conclusions}

Following recent progress in the phase-delay astrometric technique
(extension to larger distances, larger sets of radio sources, and
higher frequencies) we have initiated a program to study
the 13 radio sources of the S5 polar cap sample.   We
observed the sources twice at 8.4\,GHz and twice at 15\,GHz,
and new observations are planned at these frequencies and
at 43\,GHz.  The astrometric data reduction includes
the modelling of the atmospheric, ionospheric and source-structure
effects.  
Our first results show the feasibility of phase-connecting
the 13 radio sources at 8.4\,GHz, and obtain precisions in 
the relative separation determinations better than 0.1\,mas.  
The absolute kinematics of the radio sources will thus 
be determined unambiguously.  Furthermore,
using the astrometric information available at 8.4, 15 and 43\,GHz,
it will be possible for the first time to make rigorous spectral-index
maps, together with a multiwavelength study of the absolute kinematics of 
the S5 polar cap sources.
This observing scheme could be extended in the future to further regions
of the sky, on-route towards global phase-delay astrometry.

\section*{References}
\begin{multicols}{2}
$\bullet$ Bartel \et\ {\em Nature}, {\bf 319}, 733 (1986) \\
$\bullet$ Blandford \& K\"onigl {\em ApJ}, {\bf 232}, 34 (1979) \\
$\bullet$ Eckart \et\ {\em A\&A}, {\bf 168}, 17 (1986) \\
$\bullet$ Fey \et\ {\em ApJS}, {\bf 105}, 299 (1996) \\
$\bullet$ Fey \& Charlot {\em ApJS}, {\bf 111}, 95 (1997) \\
$\bullet$ Fey \& Charlot {\em ApJS}, {\bf 128}, 17 (2000) \\
$\bullet$ G\'omez \et\ {\em ApJ}, {\bf 435}, L19 (1995) \\
$\bullet$ Guirado \et\ {\em A\&A}, {\bf 293}, 613 (1995a) \\
$\bullet$ Guirado \et\ {\em AJ}, {\bf 110}, 2586 (1995b) \\
$\bullet$ Guirado \et\ {\em A\&A}, {\bf 336}, 385 (1998) \\
$\bullet$ Guirado \et\ {\em A\&A}, {\bf 353}, L37 (2000)  \\
$\bullet$ Johnston \& de Vegt {\em ARAA}, {\bf 37}, 97 (1999) \\
$\bullet$ K\"uhr \et\ {\em A\&ASS}, {\bf 45}, 367 (1981)  \\
$\bullet$ Lara \et\ {\em A\&A}, {\bf 314}, 672 (1996) \\
$\bullet$ Marcaide \et\ {\em AJ}, {\bf 88}, 1183 (1983) \\
$\bullet$ Marcaide \et\ {\em AJ}, {\bf 108}, 368 (1994) \\
$\bullet$ P\'erez-Torres \et\ {\em A\&A}, {\bf 360}, 161 (2000) \\
$\bullet$ Rioja \et\ {\em A\&A}, {\bf 325}, 383 (1997) \\
$\bullet$ Ros \et\ {\em A\&A}, {\bf 348}, 381 (1999) \\
$\bullet$ Ros \et\ {\em A\&A}, {\bf 356}, 357 (2000a) \\
$\bullet$ Ros \et\ {\em Proc.\ of 5th EVN VLBI Symp.}, Polatidis A.,
Conway J., Booth R., (eds.), Onsala Space Observatory, in press (2000b)\\
$\bullet$ Shapiro \et\ {\em AJ}, 84, 1459 (1979)
\end{multicols}

\end{document}